\documentstyle [12pt] {article}
\begin{document}
\setcounter{page}{1}
\def\theequation{\arabic{section}.\arabic{equation}}
\def\theequation{\thesection.\arabic{equation}}
\setcounter{section}{0}

\title{Confinement potential \\ in dual Monopole Nambu--Jona--Lasinio model
with dual Dirac strings\thanks{Supported by the Fonds zur F\"orderung der
wissenschaftlichen Forschung, Project P12495-TPH.}}

\author{M. Faber\thanks{E--mail: faber@kph.tuwien.ac.at, Tel.:
+43--1--58801--14261, Fax: +43--1--5864203} ,  A. N. Ivanov\thanks{E--mail:
ivanov@kph.tuwien.ac.at, Tel.: +43--1--58801--14261, Fax:
+43--1--5864203}~{\large{$^{\P}$}} , A. M\"uller\thanks{E--mail:
mueller@kph.tuwien.ac.at, Tel.: +43--1--58801--14265, Fax: +43--1--5864203}
,
 \\ N. I. Troitskaya\thanks{Permanent Address:
State Technical University, Department of Nuclear
Physics, 195251 St. Petersburg, Russian Federation} ~and~  M.
Zach\thanks{E--mail: zach@kph.tuwien.ac.at, Tel.: +43--1--58801--14266,
Fax: +43--1--5864203}}

\date{}

\maketitle

\begin{center}
{\it Institut f\"ur Kernphysik, Technische Universit\"at Wien, \\
Wiedner Hauptstr. 8-10, A-1040 Vienna, Austria}
\end{center}

\vskip1.0truecm
\begin{center}
\begin{abstract}
Interquark confinement potential is calculated in the dual Monopole
Nambu--Jona--Lasinio model with dual Dirac strings suggested in Refs.[1,2]
as a functional of a dual Dirac string length. The calculation is carried
out by the explicit integration over quantum fluctuations of a dual--vector
field (monopole--antimonopole collective excitation) around the  Abrikosov
flux line and string shape fluctuations. The contribution of the scalar
field (monopole--antimonopole collective excitation) exchange is taken into
account
in the tree approximation due to the London limit regime.
\end{abstract}
\end{center}

\newpage

\section{Introduction}
\setcounter{equation}{0}

The dual Monopole Nambu--Jona--Lasinio model (MNJL) with dual Dirac strings
as continuum space--time analogy of Compact Quantum Electrodynamics (CQED)
[1] has been formulated in Ref.[2--4]. The MNJL--model is based on a
Lagrangian, invariant under  magnetic $U(1)$ symmetry, with massless
magnetic monopoles  self--coupled through a local four--monopole
interaction [2,3]:
\begin{eqnarray}\label{label1.1}
\hspace{-0.3in}{\cal L}(x) = \bar{\chi}(x) i \gamma^{\mu}
 \partial_{\mu} \chi(x) + G [\bar{\chi}(x) \chi(x)]^2 - G_1
[\bar{\chi}(x)\gamma_{\mu}\chi(x)][\bar{\chi}(x)
 \gamma^{\mu}\chi(x)],
\end{eqnarray}
where $\chi(x)$ is a massless magnetic monopole field, $G$ and $G_1$ are
positive phenomenological constants. Below we show that  $G_1 = G/4$. The
magnetic monopole condensation accompanies itself the creation of massive
magnetic monopoles $\chi_{M}(x)$ with  mass $M$,
$\bar{\chi}\chi$--collective excitations with quantum numbers of the scalar
Higgs field $\sigma$ with the mass $M_{\sigma} = 2\,M$ and the massive
dual--vector field $C_{\mu}$ with the mass $M_C$ defined as [2,3]:
\begin{eqnarray}\label{label1.2}
M^2_C = \frac{g^2}{2 G_1}-\frac{g^2}{8\pi^2}[J_1(M) + M^2 J_2(M)],
\end{eqnarray}
where $J_1(M)$ and $J_2(M)$ are quadratically and logarithmically divergent
integrals [1,2]
\begin{eqnarray}\label{label1.3}
\hspace{-0.2in}J_1(M)&=&\int\frac{d^4k}{\pi^2i}\frac{1}{M^2 - k^2} =
 \Lambda^2 - M^2{\ell n}\Bigg(1
 + \frac{\Lambda^2}{M^2}\Bigg),\nonumber\\
\hspace{-0.2in}J_2(M)&=&\int \frac{d^4k}{\pi^2i}
 \frac{1}{(M^2 - k^2)^2} = {\ell n}\Bigg(1
+ \frac{\Lambda^2}{M^2}\Bigg)
- \frac{\Lambda^2}{M^2 + \Lambda^2}.
\end{eqnarray}
Here $\Lambda$ is the ultra--violet cut--off.
The mass of the massive magnetic monopole field $\chi_M(x)$ obeys the
gap--equation [2,3]:
\begin{eqnarray}\label{label1.4}
M = -2 G <\bar{\chi}(0) \chi(0)> =\frac{GM}{2\pi^2} J_1(M)
\end{eqnarray}
derived from the effective Lagrangian of the scalar $\sigma(x)$ and the
dual--vector $C_{\mu}(x)$ fields by virtue of the suppression of the direct
transitions $\sigma \longleftrightarrow {\it vacuum}$. On the other hand,
due to one--loop corrections to the mass of the monopole field derived by
using  the Lagrangian Eq.(\ref{label1.1}) the gap--equation should read
\begin{eqnarray}\label{label1.5}
M = -2\Bigg(\frac{3}{4}\,G + G_1\Bigg) <\bar{\chi}(0) \chi(0)>.
\end{eqnarray}
Since the level of the collective $\bar{\chi}\chi$--excitations should be
completely compatible with the monopole level, the gap--equations
Eq.(\ref{label1.2}) and Eq.(\ref{label1.4}) should coincide. This fixes
$G_1$ in terms of $G$ as $G_1 = G/4$.

As has been shown in Refs.[2, 3] the vacuum expectation values of time
ordered products of densities expressed in terms of the massless monopole
field, i.e., the magnetic monopole Green functions
\begin{eqnarray}\label{label1.6}
G\,(x_{1},\ldots, x_{n})=<{0}|{\rm T}(\bar{\chi} (x_{1})
 \Gamma_{1} \chi (x_{1}) \ldots \bar{\chi} (x_{n})
\,\Gamma_{n} \chi\,(x_{n}))|{0}>_{\rm conn.}\,,
\end{eqnarray}
where $\Gamma_i (i = 1,\ldots,n)$ are the Dirac matrices, are given by [2,3]
\begin{eqnarray}\label{label1.7}
\hspace{-0.3in}&&G(x_{1},\ldots,x_{n})=<0|{\rm T}(\bar{\chi}(x_{1})
 \Gamma_1 \chi(x_{1}) \ldots \bar{\chi}(x_{n})
\Gamma_n \chi(x_{n}))|{0}>_{\rm conn.} = \nonumber\\
\hspace{-0.3in}&&= ^{(M)}\!<{0}|{\rm T} \Big( \bar{\chi}_M(x_{1})
 \Gamma_{1} \chi_M(x_{1}) \ldots \bar{\chi}_M(x_{n}) \Gamma_{n}
\chi_M(x_{n})\nonumber\\
\hspace{-0.3in}&&\times\,\exp i\int d^4x \{-g\bar{\chi}_M(x)
\gamma^\nu \chi_M(x) C_{\nu}(x)- \kappa \bar{\chi}_M(x) \chi_M(x)
 \sigma(x) \nonumber\\
\hspace{-0.3in}&&+ {\cal L}_{\rm int}
 [\sigma(x)]\}\Big)|0>^{(M)}_{\rm conn.}.
\end{eqnarray}
Here $|{0}>^{(M)}$ is the wave function of the non--perturbative vacuum of
the MNJL--model in the condensed phase and $|{0}>$ the wave function of the
perturbative vacuum of the non--condensed phase. Then, ${\cal L}_{\rm
int}[\sigma(x)]$ describes self--interactions of the $\sigma$--field:
\begin{eqnarray}\label{label1.8}
{\cal L}_{\rm int}[\sigma(x)] = - \kappa\,M_{\sigma}\,\sigma^3(x) -
\frac{1}{2}\,\sigma^4(x).
\end{eqnarray}
The self--interactions ${\cal L}_{\rm int}[\sigma(x)]$ provide
$\sigma$--field loop contributions and can be dropped out in the tree
$\sigma$--field approximation accepted in Refs. [2--4]. The tree
$\sigma$--field approximation can be justified keeping massive magnetic
monopoles very heavy, i.e.
$M \gg M_C$. This corresponds to the London limit $M_{\sigma} =
2\,M \gg M_C$ in the dual Higgs model with dual Dirac strings [5--7]. The
inequality
$M_{\sigma} \gg M_C$ means also that in the MNJL model we deal with
{\it Dual Superconductivity of type {\rm I$\!$I}} [4]. In the tree
$\sigma$--field approximation the r.h.s. of Eq.(\ref{label1.7}) acquires
the form
\begin{eqnarray}\label{label1.9}
\hspace{-0.3in}&&G(x_{1},\ldots,x_{n})=<{0}|{\rm T}(\bar{\chi}(x_{1})
 \Gamma_1 \chi(x_{1}) \ldots \bar{\chi}(x_{n})
 \Gamma_n \chi(x_{n}))|{0}>_{\rm conn.} = \nonumber\\
\hspace{-0.3in}&&= ^{(M)}\!<{0}|{\rm T}
 \Big( \bar{\chi}_M(x_{1}) \Gamma_{1} \chi_M(x_{1})
 \ldots \bar{\chi}_M(x_{n}) \Gamma_{n} \chi_M(x_{n})\nonumber\\
\hspace{-0.3in}&&\exp i\int d^4x \Big\{-g\bar{\chi}_M(x)
 \gamma^\nu \chi_M(x) C_{\nu}(x) - \kappa \bar{\chi}_M(x)
 \chi_M(x) \sigma(x)\}\Big)|{0}>^{(M)}_{\rm conn.}.
\end{eqnarray}
For the subsequent investigation it is convenient to represent the r.h.s.
of Eq.(\ref{label1.9}) in terms of the generating functional of the
monopole Green functions [2--4]
\begin{eqnarray}\label{label1.10}
\hspace{-0.3in}&&G(x_{1},\ldots,x_{n}) =
 \prod^{n}_{i=1}\frac{\delta}{\delta \eta(x_i)}
 \Gamma_i \frac{\delta}{\delta \bar{\eta}(x_i)}
 Z[\eta,\bar{\eta}]\Bigg|_{\eta =\bar{\eta}=0},
\end{eqnarray}
where $\bar{\eta}(\eta)$ are the external sources of the massive monopole
(antimonopole) fields, and $Z[\eta,\bar{\eta}]$ is the generating
functional of the monopole Green functions defined by
\begin{eqnarray}\label{label1.11}
\hspace{-0.3in}&&Z[\eta,\bar{\eta}]=\frac{1}{Z}\int
{\cal D}\chi_M {\cal D}\bar{\chi}_M {\cal D}C_{\mu}
{\cal D}\sigma\,\exp i\int d^4x
\,\Big[ \frac{1}{4}\,F_{\mu\nu}(x)\,F^{\mu\nu}(x)\nonumber\\
\hspace{-0.3in}&&+ \frac{1}{2}\,M^2_C\,C_{\mu}(x)\,C^{\mu}(x)
+ \frac{1}{2}\,\partial_{\mu}\sigma(x)\,\partial^{\mu} \sigma(x)
 - \frac{1}{2}\,M^2_{\sigma}\,\sigma^2(x)\nonumber\\
\hspace{-0.3in}&&+ \bar{\chi}_M(x)(i\,\gamma^{\mu}\,\partial_{\mu}
- M - g\,\gamma^{\mu}\,C_{\mu}(x)
- \kappa\,\sigma(x))\,\chi_M(x)\nonumber\\
\hspace{-0.5in}&&+ \bar{\eta}(x)\,\chi_M(x)
+ \bar{\chi}_M(x)\,\eta(x) + {\cal L}_{\rm free~quark}(x)\Big].
\end{eqnarray}
The normalization factor $Z$ is defined by the condition $Z[0,0] = 1$. The
coupling constants $g$ and $\kappa$ are related by the constraint
\begin{eqnarray}\label{label1.12}
\frac{g^2}{12\pi^2}J_2(M)=\frac{\kappa^2}{8\pi^2}\,J_2(M) = 1
\end{eqnarray}
or $\kappa^2=2g^2/3$ [2,3]. Then, ${\cal L}_{\rm free~quark}(x)$ is the
kinetic term for the quark and antiquark
\begin{eqnarray}\label{label1.13}
\hspace{-0.3in}{\cal L}_{\rm free~quark}(x) =
 - \sum_{i=q,\bar{q}} m_i \int d{\tau}
 \Bigg(\frac{d X^{\mu}_i(\tau)}{d\tau}
 \frac{d X^{\nu}_i(\tau)}{d\tau} g_{\mu\nu}\Bigg)^{1/2}
 \delta^{(4)} (x - X_i(\tau)).
\end{eqnarray}
In our consideration quarks and antiquarks are classical point--like
particles with masses $m_q = m_{\bar{q}} = m$, electric charges
 $Q_q = - Q_{\bar{q}} = Q$, and trajectories $X^{\nu}_q(\tau)$ and
$X^{\nu}_{\bar{q}}(\tau)$,  respectively. The field strength
$F^{\mu\nu}(x)$ is defined [2--4] as
$F^{\mu\nu}(x) = {\cal E}^{\mu\nu}(x) - {^*}dC^{\mu\nu}(x)$, where
$dC^{\mu\nu}(x) = \partial^{\mu} C^{\nu} (x)
 - \partial^{\nu} C^{\mu}(x)$, and ${^*}dC^{\mu\nu}(x)$ is the dual
version, i.e.,
${^*}dC^{\mu\nu}(x) = \frac{1}{2} \varepsilon^{\mu\nu\alpha\beta}
dC_{\alpha\beta}(x)\,(\varepsilon^{0123} = 1)$. The dual "chromo"--electric
field strength  ${\cal E}^{\mu\nu}(x)$, induced by a dual Dirac string, is
defined following [2--4] as
\begin{eqnarray}\label{label1.14}
\hspace{-0.2in}&&{\cal E}^{\mu\nu}(x) = Q \int\!\!\!\int
  d\tau d \sigma \Bigg(\frac{\partial X^{\mu}}{\partial \tau}
 \frac{\partial X^{\nu}}{\partial \sigma}
 - \frac{\partial X^{\nu}}{\partial \tau}
 \frac{\partial X^{\mu}}{\partial \sigma}\Bigg)\delta^{(4)}(x - X),
\end{eqnarray}
where $X^{\mu} = X^{\mu} (\tau,\sigma)$ represents the position of a point
on the world sheet swept by the string. The sheet is parameterized by
internal coordinates  ${-\infty} < \tau < {\infty}$ and $0 \le \sigma \le
\pi$, so that $X^{\mu} (\tau, 0) = X^{\mu}_{ - Q}(\tau)$ and
$X^{\mu}(\tau,\pi) = X^{\mu}_Q(\tau)$ represent the world lines of an
anti--quark and a quark [2--7]. Within the definition Eq.(\ref{label1.14})
the tensor field ${\cal E}^{\mu\nu}(x)$ satisfies identically the equation
of motion,
  $\partial_{\mu} F^{\mu\nu}(x) = J^{\nu}(x)$. The electric quark current
$J^{\nu}(x)$ is defined as
\begin{eqnarray}\label{label1.15}
\hspace{-0.2in}J^{\nu}(x) = \sum_{i= q,\bar{q}} Q_i\int d \tau \frac{d
X^{\nu}_i(\tau)}{d\tau} \delta^{(4)}(x  - X_i(\tau)).
\end{eqnarray}
Hence, the inclusion of a dual Dirac string in terms of
 ${\cal E}^{\mu\nu}(x)$ defined by Eq.(\ref{label1.14}) satisfies
completely the dual electric Gauss law of Dirac$^{\prime}$s extension of
 Maxwell$^{\prime}\,$s electrodynamics.

The ground state of the massive dual--vector field $C_{\mu}(x)$ coupled to
a dual Dirac string acquires the shape of the  Abrikosov flux line [2--7]
\begin{eqnarray}\label{label1.16}
C^{\nu}[{\cal E}(x)] = - \int d^4x^{\prime}\,\Delta(x -
x^{\prime}\,)\,\partial_{\mu}{^*{\cal E}^{\mu\nu}(x^{\prime}\,)},
\end{eqnarray}
where $\Delta(x - x^{\prime}\,)$ is the Green function
\begin{eqnarray}\label{label1.17}
\Delta(x - x^{\prime}\,) = \int \frac{d^4k}{(2\pi)^4}
\,\frac{e^{\displaystyle  - i k\cdot (x
- x^{\prime}\,)}}{M^2_C - k^2 - i 0}.
\end{eqnarray}
Integrating out the dual--vector field fluctuations $c_{\mu}(x)$ around
the shape of the  Abrikosov flux line, $C_{\mu}(x) = C_{\mu}[{\cal E}(x)] +
c_{\mu}(x)$, and the scalar $\sigma$--field [4] we obtain the generating
functional of the monopole Green functions in the following form:
\begin{eqnarray}\label{label1.18}
\hspace{-0.3in}&&Z[\eta,\bar{\eta}]=\frac{1}{Z}
 \int {\cal D}\chi_M {\cal D}\bar{\chi}_M \,\exp i\int d^4x
\,\Big[{\cal L}_{\rm eff}\{\bar{\chi}_M(x), \chi_M(x),
 C^{\nu}[{\cal E}(x)]\}\nonumber\\
\hspace{-0.3in}&&+ \bar{\chi}_M(x)(i\,\gamma^{\mu}\,\partial_{\mu}
- M - g\,\gamma^{\mu}\,C_{\mu}[{\cal E}(x)])\,\chi_M(x)
 + \bar{\eta}(x)\,  \chi_M(x)\nonumber\\
\hspace{-0.3in}&&+ \bar{\chi}_M(x)\,\eta(x)
 + {\cal L}_{\rm free~quark}(x)\Big],
\end{eqnarray}
where ${\cal L}_{\rm eff}\{\bar{\chi}_M(x), \chi_M(x),
 C^{\nu}[{\cal E}(x)]\}$ reads
\begin{eqnarray}\label{label1.19}
\hspace{-0.3in}&&{\cal L}_{\rm eff}\{\bar{\chi}_M(x), \chi_M(x),
 C^{\nu}[{\cal E}(x)]\} =
 {\cal L}_{\rm string}\{C^{\nu}[{\cal E}(x)]\}\nonumber\\
\hspace{-0.3in}&& - \frac{g^2 }{2M^2_C}[\bar{\chi}_M(x) \gamma_{\mu}
\chi_M(x)]\,[\bar{\chi}_M(x) \gamma^{\mu} \chi_M(x)]
+ \frac{\kappa^2 }{2M^2_{\sigma}}[\bar{\chi}_M(x) \chi_M(x)]^2.
\end{eqnarray}
The Lagrangian of the dual Dirac string
${\cal L}_{\rm string}\{C^{\nu}[{\cal E}(x)]\}$ is defined [3--6]
\begin{eqnarray}\label{label1.20}
\hspace{-0.5in}\int d^4x
\,{\cal L}_{\rm string}\{C^{\nu}[{\cal E}(x)]\} = \frac{1}{4}
\,M^2_C\int \int d^4x d^4y {\cal E}_{\mu\alpha}(x)
\,\Delta^{\alpha}_{\nu}(x - y, M_C){\cal E}^{\mu\nu}(y),
\end{eqnarray}
where $\Delta^{\alpha}_{\nu}(x - y, M_C) = (g^{\alpha}_{\nu} +
2\partial^{\alpha}\partial_{\nu}/M^2_C) \Delta (x - y; M_C)$.

The effective Lagrangian Eq.(\ref{label1.19}) integrated over the massive
monopole fields $\bar{\chi}_M(x)$ and $\chi_M(x)$ defines the string
energy, i.e. the interquark potential, as a functional of the string shape.

\section{Confinement potential}
\setcounter{equation}{0}

The interquark confinement potential is related to the energy of the string
which is defined as follows [2--7]:
\begin{eqnarray}\label{label2.1}
\hspace{-0.3in}W &=& - \int d^3x
\,{\cal L}_{\rm string}\{C^{\nu}[{\cal E}(x)]\}
 +  \int d^3x \,{^{(M)}<0|}{\rm T}\Big(\Big(- \frac{g^2
}{2M^2_C}[\bar{\chi}_M(x) \gamma_{\mu} \chi_M(x)]\nonumber\\
\hspace{-0.3in}&& \times\,[\bar{\chi}_M(x)
 \gamma^{\mu} \chi_M(x)]
+ \frac{\kappa^2 }{2M^2_{\sigma}}
[\bar{\chi}_M(x) \chi_M(x)]^2\Big)\nonumber\\
\hspace{-0.3in}&&\times
\,\exp -ig \int d^4y\,\bar{\chi}_M(y)\,\gamma^{\mu}
\,C_{\mu}[{\cal E}(y)]\,\chi_M(y) \Big){|0>^{(M)}}.
\end{eqnarray}
The interaction caused by the integration over the $\sigma$--field
fluctuations gives a trivial constant contribution to the energy of the
string [4] and can be dropped out. In the momentum representation of the
vacuum expectation values the energy of the string is then defined by [4]:
\begin{eqnarray}\label{label2.2}
\hspace{-0.2in}&&W = - \int d^3x
\,{\cal L}_{\rm string}\{C^{\nu}[{\cal E}(x)]\}
- \int d^3x \,\frac{g^2}{2M^2_C}
\,\int \frac{d^4k_1}{(2\pi)^4i}\,\nonumber\\
\hspace{-0.2in}&&\times\,{\rm tr}
 \Bigg\{\frac{1}{M - \hat{k}_1 + g \hat{C}[{\cal E}(x)]}
 \gamma^{\mu}\Bigg\}\int \frac{d^4k_2}{(2\pi)^4i}
\,{\rm tr}\Bigg\{\gamma_{\mu}\frac{1}{M - \hat{k}_2
 + g \hat{C}[{\cal E}(x)]}\Bigg\}.
\end{eqnarray}
The momentum integrals have been calculated in Ref.[4]. This yields the
energy of the string:
\begin{eqnarray}\label{label2.3}
\hspace{-0.2in} W &=& - \int d^3x
\,{\cal L}_{\rm string}\{C^{\nu}[{\cal E}(x)]\} \nonumber\\
\hspace{-0.2in}&& - \frac{1}{2}\,\frac{1}{M^2_C}
\,\Bigg(\frac{g^2}{8\pi^2}[J_1(M) + M^2 J_2(M)]\Bigg)^2
 \int d^3x \,C_{\mu}[{\cal E}(x)]\,C^{\mu}[{\cal E}(x)].
\end{eqnarray}
By using Eqs.(\ref{label1.2}) -- (\ref{label1.4}), the relations $G_1 =
G/4$ and $M_{\sigma} = 2\,M$ we bring up the coefficient of the second term
to the form
\begin{eqnarray}\label{label2.4}
- \frac{g^2}{8\pi^2}[J_1(M) + M^2 J_2(M)] = M^2_C
 - \frac{g^2}{2G_1}= M^2_C
 + 8\,g^2 \frac{<\bar{\chi}\chi>}{M_{\sigma}}.
\end{eqnarray}
Thus, the energy of the string containing quantum fluctuations of the
scalar and dual--vector fields around the shape of the Abrikosov flux line
is given by
\begin{eqnarray}\label{label2.5}
\hspace{-0.2in}&&W = - \int d^3x
\,{\cal L}_{\rm string}\{C^{\nu}[{\cal E}(x)]\}\nonumber\\
\hspace{-0.2in}&&- \frac{1}{2}\,M^2_C
\,\Bigg(1 + \frac{8 g^2}{M^2_C}
\,\frac{<\bar{\chi}\chi>}{M_{\sigma}}\Bigg)^2\int d^3x
\,C_{\mu}[{\cal E}(x)]\,C^{\mu}[{\cal E}(x)].
\end{eqnarray}
The computation of the r.h.s. of Eq.(\ref{label2.5}) we perform for the
static straight string of the length $L$  directed along the $z$--axis. In
this case the electric field strength ${\cal E}_{\mu\nu}(x)$ does not
depend on time and is given by [6]
\begin{eqnarray}\label{label2.6}
\vec{\cal E}(\vec{x}\,) = \vec{e}_z\,Q\,\delta (x)\,\delta (y)
\,\Bigg[\theta \Bigg(z -\frac{1}{2}\,L\Bigg)
 - \theta \Bigg(z +\frac{1}{2}\,L\Bigg)\Bigg],
\end{eqnarray}
where a quark and an antiquark are placed at
$\vec{X}_Q = (0,0,\frac{1}{2}\,L)$ and
$\vec{X}_{ - Q} = (0,0,-\frac{1}{2}\,L)$. The unit vector $\vec{e}_z$ is
directed along the $z$--axis and $\theta (z)$ is the step--function. The
field strength Eq.(\ref{label2.6}) induces the dual--vector potential
\begin{eqnarray}\label{label2.7}
<\vec{C}(\vec{x}\,)> = - \,i\,Q\,\int \frac{d^3k}{4\,\pi^3}
\,\frac{\vec{k} \times \vec{e}_z}{k_z}\,\frac{1}{M^2_C +
\vec{k}^{\,2}}\,\sin\Bigg(\frac{k_z L}{2}\Bigg)
\, e^{\displaystyle i\,\vec{k}\cdot\vec{x}}.
\end{eqnarray}
For the static straight string the term
$-\int d^3x{\cal L}_{\rm string}\{C^{\nu}[{\cal E}(x)]\}$ reads
\begin{eqnarray}\label{label2.8}
 \hspace{-0.3in}&&-\int d^3x{\cal L}_{\rm string}\{C^{\nu}[{\cal E}(x)]\} =
-\frac{1}{4}\,M^2_C \int d^3x \int d^3x^{\prime}
 \int\limits^{\infty}_{-\infty} d x^{\prime}_0 \nonumber\\
 \hspace{-0.3in}&&\times\,\Bigg[{\cal E}_{0 i}(\vec{x}\,)
\,\Bigg(g^i_j + \frac{2}{M^2_C}
\,\frac{\partial^2}{\partial x_i \partial x^j}\Bigg)
\,\Delta (x_0 - x^{\prime}_0, \vec{x} - \vec{x}^{\,\prime}, M_C)
\,{\cal E}^{0j}(\vec{x}^{\,\prime}\,) \nonumber\\
 \hspace{-0.3in}&&+ {\cal E}_{i 0}(\vec{x}\,)\Bigg(g^0_0 +
\frac{2}{M^2_C}\frac{\partial^2}{\partial x_0 \partial x^0}\Bigg)
\,\Delta (x_0 - x^{\prime}_0, \vec{x} - \vec{x}^{\,\prime}, M_C)
\,{\cal E}^{i0}(\vec{x}^{\,\prime}\,)\Bigg] =\nonumber\\
\hspace{-0.3in}&&= \frac{1}{2}\,M^2_C \int d^3x \int d^3x^{\prime}
\,\vec{{\cal E}}(\vec{x}\,)\cdot \vec{{\cal E}}(\vec{x}^{\,\prime}\,)
\,\Bigg(1 - \frac{1}{M^2_C}\frac{\partial^2}{\partial z^2}\Bigg)
\,\Delta(\vec{x} - \vec{x}^{\,\prime}, M_C) =\nonumber\\
\hspace{-0.3in}&&= \frac{1}{2}\,Q^2 M^2_C \int\limits^{L/2}_{-L/2}
dz\int\limits^{L/2}_{-L/2} d z^{\prime}
 \int\limits^{\infty}_{-\infty} d k_z\,\Bigg(1 +
 \frac{k^2_z}{M^2_C}\Bigg)\int\frac{d^2k_{\perp}}{(2\pi)^3}\,
  \frac{e^{\displaystyle ik_z(z - z^{\prime}\,)}}
{M^2_C + \vec{k}^{\,2}_{\perp} + k^2_z}=\nonumber\\
 \hspace{-0.3in}&&= \frac{Q^2 M^2_C}{4\pi^3}
 \int\limits^{\infty}_{-\infty}\frac{dk_z}{k^2_z}\,
 \sin^2\Bigg(\frac{k_zL}{2}\Bigg)
\,\Bigg(1 + \frac{k^2_z}{M^2_C}\Bigg)
 \int\frac{d^2k_{\perp}}{M^2_C + \vec{k}^{\,2}_{\perp} + k^2_z}
 =\nonumber\\
 \hspace{-0.3in}&&= \frac{Q^2 M^2_C}{4\pi^2}
 \int\limits^{\infty}_{-\infty}\frac{dk_z}{k^2_z}\,
 \sin^2\Bigg(\frac{k_zL}{2}\Bigg)
\,\Bigg(1 + \frac{k^2_z}{M^2_C}\Bigg)
 \int\limits^{\Lambda^2_{\perp}}_0
 \frac{dk^2_{\perp}}{M^2_C + k^2_{\perp} + k^2_z},
\end{eqnarray}
where $\Lambda_{\perp}$ is the cut--off in the plane perpendicular to the
world--sheet swept by the string [2--7]. We identify $\Lambda_{\perp}$ with
the mass of the scalar field, i.e., $\Lambda_{\perp} = M_{\sigma} = 2 M$
[2--7].

For a sufficiently long string we can integrate over $k^2_{\perp}$ and get
\begin{eqnarray}\label{label2.9}
 \hspace{-0.2in}-\int d^3x
\,{\cal L}_{\rm string}\{C^{\nu}[{\cal E}(x)]\} &=&
 \frac{Q^2 M^2_C}{4\pi^2} \int\limits^{\infty}_{-\infty}
 \frac{dk_z}{k^2_z}\,\sin^2\Bigg(\frac{k_zL}{2}\Bigg)
 \,\Bigg(1 + \frac{k^2_z}{M^2_C}\Bigg)\nonumber\\
&&\times\,\Bigg[ {\ell n}\Bigg(1 + \frac{M^2_{\sigma}}{M^2_C}\Bigg)
 - {\ell n}\Bigg(1 + \frac{k^2_z}{M^2_C}\Bigg)\Bigg],
\end{eqnarray}
where we have neglected $k_z$ relative to $\Lambda_{\perp}$. Dropping the
infinite constant  contributions independent of $L$ we obtain [6]:
\begin{eqnarray}\label{label2.10}
\hspace{-0.3in}-\int d^3x
\,{\cal L}_{\rm string}\{C^{\nu}[{\cal E}(x)]\} &=&
 L \frac{Q^2 M^2_C}{8\pi} \Bigg[{\ell n}\Bigg(1 +
\frac{M^2_{\sigma}}{M^2_C}\Bigg) + 2\,E_1(M_CL) \nonumber\\
\hspace{-0.3in}&&- \frac{2}{M_CL}
\,\Bigg(1 - e^{\displaystyle - M_C L}\Bigg) \Bigg]
 - \frac{Q^2}{4\pi}\,\frac{e^{\displaystyle - M_C L}}{L},
\end{eqnarray}
where $E_1(M_CL)$ is the Exponential Integral function. For the calculation
of the integral over $k_z$ we have used the auxiliary integral
\begin{eqnarray}\label{label2.11}
\hspace{-0.4in}&&\int\limits^{\infty}_{-\infty}dx
\,\frac{\sin^2x}{x^2}
\,{\ell n}\Bigg(\alpha^2 + \frac{x^2}{a^2}\Bigg) = 2\pi
\,{\ell n}\alpha + \frac{\pi}{a \alpha}\Bigg(1
 - e^{\displaystyle - 2 a \alpha}\Bigg) - 2\pi\,E_1(2 a\alpha),
\end{eqnarray}
where in Eq.(\ref{label2.10}) we have set $\alpha = 1$ and $a = M_C L/2$.

The first term proportional to $L$ gives the string tension $\sigma_0$
calculated in the tree--approximation [5]:
\begin{eqnarray}\label{label2.12}
\sigma_0 = \frac{Q^2 M^2_C}{8\pi} {\ell n}\Bigg(1 +
\frac{M^2_{\sigma}}{M^2_C}\Bigg).
\end{eqnarray}
The last term in Eq.(\ref{label2.5}) induced by the quantum fluctuations of
the dual--vector field $C_{\mu}$ around the shape of the  Abrikosov flux
line can be reduced to the form [6]:
\begin{eqnarray}\label{label2.13}
 \hspace{-0.4in}&& - \frac{1}{2}
\,M^2_C\Bigg(1 + \frac{8 g^2}{M^2_C}
\frac{<\bar{\chi}\chi>}{M_{\sigma}}\Bigg)^2 \int d^3x
 \,C_{\mu}[{\cal E}(x)]\,C^{\mu}[{\cal E}(x)] =\nonumber\\
\hspace{-0.4in}&&= \frac{Q^2 M^2_C}{4\pi^2}
 \,\Bigg(1 + \frac{8 g^2}{M^2_C}
 \frac{<\bar{\chi}\chi>}{M_{\sigma}}\Bigg)^2
 \int\limits^{\infty}_{-\infty}\frac{dk_z}{k^2_z}\,
  \sin^2\Bigg(\frac{k_zL}{2}\Bigg)
\,\Bigg[{\ell n}\Bigg(1
 + \frac{M^2_{\sigma}}{M^2_C}\Bigg)
 \nonumber\\
 \hspace{-0.4in}&&- {\ell n}\Bigg(1
+ \frac{k^2_z}{M^2_C}\Bigg)\Bigg] =
 L\,\frac{Q^2 M^2_C}{8\pi}\,\Bigg(1
 + \frac{8 g^2}{M^2_C}
  \frac{<\bar{\chi}\chi>}{M_{\sigma}}\Bigg)^2
  \nonumber\\
&&\times\,\Bigg[{\ell n}\Bigg(1
 + \frac{M^2_{\sigma}}{M^2_C}\Bigg)
+ 2\,E_1(M_CL)
- \frac{2}{M_CL}
\,\Bigg(1 - e^{\displaystyle - M_C L}\Bigg)\Bigg].
\end{eqnarray}
Collecting the pieces together we obtain the energy of the dual Dirac
string, the interquark potential, as a function of the the length of the
string $L$:
\begin{eqnarray}\label{label2.14}
 \hspace{-0.2in}&&W = L\,\frac{Q^2 M^2_C}{4\pi}\,
 \Bigg(1 + \frac{8 g^2}{M^2_C}\frac{<\bar{\chi}\chi>}{M_{\sigma}}
 + \frac{32 g^4}{M^4_C}\frac{<\bar{\chi}\chi>^2}{M^2_{\sigma}}\Bigg)
  \Bigg[{\ell n}\Bigg(1
 + \frac{M^2_{\sigma}}{M^2_C}\Bigg) \nonumber\\
 \hspace{-0.2in}&&+ 2\,E_1(M_CL) -\frac{2}{M_CL}
\,\Bigg(1 - e^{- M_C L}\Bigg)\Bigg]
- \frac{Q^2}{4\pi}\,\frac{e^{\displaystyle - M_C L}}{L}.
\end{eqnarray}
The term proportional to $L$ describes a linearly rising interquark
potential leading to the quark confinement and gives the expression for the
string tension
\begin{eqnarray}\label{label2.15}
 \sigma =\frac{Q^2 M^2_C}{4\pi}\,\Bigg(1
+ \frac{8 g^2}{M^2_C}\frac{<\bar{\chi}\chi>}{M_{\sigma}}
 + \frac{32 g^4}{M^4_C}
 \frac{<\bar{\chi}\chi>^2}{M^2_{\sigma}}\Bigg)
\,{\ell n}\Bigg(1 + \frac{M^2_{\sigma}}{M^2_C}\Bigg).
\end{eqnarray}
The last term in Eq.(\ref{label2.14}) is the Yukawa potential.

Matching the string tension Eq.(\ref{label2.15}) with the string tension
$\sigma_0$ calculated in the tree--approximation Eq.(\ref{label2.12}) we
accentuate a tangible contribution of quantum fluctuations of the
dual--vector field $C_{\mu}$ around the shape of the  Abrikosov flux line.
This agrees with the result obtained in Ref.[6] in the dual Higgs model.

\section{String shape fluctuations}
\setcounter{equation}{0}

The string shape fluctuations we define as usually [8,7] by
$X^{\mu} \to X^{\mu} + \eta^{\mu}(X)$, where $\eta^{\mu}(X)$ describes
fluctuations around the fixed surface $S$ swept by the shape
$\Gamma$ and obeys the constraint $\eta^{\mu}(X)|_{\partial S} = 0$ [8,7]
at the boundary $\partial S$ of the surface $S$. The integration over the
$\eta$--field we perform around the shape of the static straight string
with the length $L$ tracing out the rectangular surface $S$ with the
time--side
$T$ [8,7]. Allowing only fluctuations in the plane perpendicular to the
string world--sheet and setting $\eta_t(t,z) = \eta_z(t,z) = 0$ [8,7],
we arrive at the fluctuation action $\delta
\,{\cal S}_{\rm N}[\eta_x,\eta_y]$ [7,4]
\begin{eqnarray}\label{label3.1}
\hspace{-0.2in}\delta\,{\cal S}_{\rm N}[\eta_x,\eta_y] = -
\frac{3Q^2\Lambda^2_{\perp}}{32\pi}\int\limits^{T/2}_{-T/2} dt
\int\limits^{L/2}_{-L/2} dz [\eta_x(t,z)\,(- \Delta)
 \,\eta_x(t,z) + (x \leftrightarrow y)],
\end{eqnarray}
coming from the term $\int d^4x
\,{\cal L}_{\rm string}\{C^{\nu}[{\cal E}(x)]\}$ defined by
Eq.(\ref{label1.20}), where $\Delta$ is the Laplace operator
in 2--dimensional space--time
\begin{eqnarray}\label{label3.2}
\Delta = - \frac{\partial^2}{\partial t^2}
+ \frac{\partial^2}{\partial z^2}.
\end{eqnarray}
The term $\int d^4x\,C_{\mu}[{\cal E}(x)]\,C^{\mu}[{\cal E}(x)]$ in
Eq.(\ref{label2.5}), induced by the quantum fluctuations of the
dual--vector $C_{\mu}$ and scalar $\sigma$ fields around the shape of the
Abrikosov flux line, does not contribute to the fluctuation action for the
case of the static straight string. In order to show this we use the
expression obtained in Ref.[4]:
\begin{eqnarray}\label{label3.3}
\hspace{-0.4in}&& \delta C_{\mu}[{\cal E}(x)]
\,C^{\mu}[{\cal E}(x)] = \nonumber\\
\hspace{-0.4in}&&\times\,Q^2\int\!\!\!\int
\frac{d^3k}{4\pi^3}\frac{d^3q}{4\pi^3}
\,\frac{k_x q_x + k_y q_y}{k_z q_z}\sin\Bigg(\frac{k_zL}{2}\Bigg) \,
\sin\Bigg(\frac{q_zL}{2}\Bigg)\,\frac{1}{M^2_C + \vec{k}^{\,2}}
\nonumber\\
\hspace{-0.4in}&&\times \; \frac{1}{M^2_C + \vec{q}^{\,2}}\,
 e^{\displaystyle i\,(\vec{k} + \vec{q})\cdot \vec{x}}\,
 \Big(e^{\displaystyle i\,[(k_x + q_x) \eta_x(t,z)
+ (k_y + q_y) \eta_y(t,z)]} - 1\Big).
\end{eqnarray}
The contribution to the fluctuation action is given by
\begin{eqnarray}\label{label3.4}
\hspace{-0.4in}&&\int d^4x\,\delta C_{\mu}[{\cal E}(x)]
\,C^{\mu}[{\cal E}(x)] = \nonumber\\
\hspace{-0.4in}&&= Q^2\int d^4x\int\!\!\!\int
\frac{d^3k}{4\pi^3}\frac{d^3q}{4\pi^3}
\,\frac{k_x q_x + k_y q_y}{k_z q_z}
\,\sin\Bigg(\frac{k_zL}{2}\Bigg)
\,\sin\Bigg(\frac{q_zL}{2}\Bigg)\,\frac{1}{M^2_C +
\vec{k}^{\,2}}\,\nonumber\\
\hspace{-0.4in}&&\times \; \frac{1}{M^2_C + \vec{q}^{\,2}}
\,e^{\displaystyle i\,(\vec{k} + \vec{q})\cdot \vec{x}}
\,\Big(e^{\displaystyle i\,[(k_x + q_x) \eta_x(t,z)
 + (k_y + q_y) \eta_y(t,z)]} - 1\Big).
\end{eqnarray}
Integrating over $x$ and $y$ we get
\begin{eqnarray}\label{label3.5}
\hspace{-0.4in}&&\int d^4x\,\delta C_{\mu}[{\cal E}(x)]
\,C^{\mu}[{\cal E}(x)] = \nonumber\\
\hspace{-0.4in}&&= Q^2\int\limits^{T/2}_{-T/2}dt \int\limits^{L/2}_{-L/2}dz
 \int\!\!\!\int \frac{d^3k}{2\pi^2}\frac{d^3q}{2\pi^2}
\,\frac{k_x q_x + k_y q_y}{k_z q_z}\sin\Bigg(\frac{k_zL}{2}\Bigg) \,
\sin\Bigg(\frac{q_zL}{2}\Bigg)
\,\frac{1}{M^2_C + \vec{k}^{\,2}} \nonumber\\
\hspace{-0.4in}&&\times \,\frac{1}{M^2_C + \vec{q}^{\,2}}\,
e^{\displaystyle i\,(k_0 + q_0)t - i(k_z + q_z)z}
\,\delta(k_x + q_x)\,\delta(k_y + q_y) \nonumber\\
\hspace{-0.4in}&&\times
\,\Big(e^{\displaystyle i\,[(k_x + q_x) \eta_x(t,z)
+ (k_y + q_y) \eta_y(t,z)]} - 1\Big) = 0.
\end{eqnarray}
Thus, Eq.(\ref{label3.1}) defines completely the fluctuation action induced
by string shape fluctuations around a static straight string with length
$L$. As has been shown in Ref.[7], the fluctuation action
Eq.(\ref{label3.1}) gives a Coulomb--like universal contribution [8] to the
energy of the string:
\begin{eqnarray}\label{label3.6}
W_{\rm string-shape} = - \frac{\alpha_{\rm string}}{L},
\end{eqnarray}
where $\alpha_{\rm string} = \pi/12$ and $\alpha_{\rm string} = \pi/3$ for
openned and closed strings, respectively.

\section{Conclusion}
\setcounter{equation}{0}

We have shown that in the MNJL model with dual Dirac strings the quantum
fluctuations of a dual--vector field $C_{\mu}$ and a scalar field $\sigma$
around the shape of the Abrikosov flux line give the interquark confinement
potential in the following form
\begin{eqnarray}\label{label4.1}
\hspace{-0.2in}&&W_{\rm tot} = L\,\frac{Q^2 M^2_C}{4\pi}\,
 \Bigg(1 + \frac{8 g^2}{M^2_C}\frac{<\bar{\chi}\chi>}{M_{\sigma}} +
 \frac{32 g^4}{M^4_C}\frac{<\bar{\chi}\chi>^2}{M^2_{\sigma}}\Bigg)
 \Bigg[{\ell n}\Bigg(1 + \frac{M^2_{\sigma}}{M^2_C}\Bigg)
 \nonumber\\
 \hspace{-0.2in}&& + 2\,E_1(M_CL) -\frac{2}{M_CL}\,
 \Bigg(1 - e^{- M_C L}\Bigg)\Bigg] - \frac{Q^2}{4\pi}\,
 \frac{e^{\displaystyle - M_C L}}{L} -
 \frac{\alpha_{\rm string}}{L},
\end{eqnarray}
where $\alpha_{\rm string} = \pi/12$ and $\alpha_{\rm string} = \pi/3$ for
openned and closed strings, respectively. This interquark potential
resembles the result obtained in the dual Higgs model with dual Dirac
strings [5--7]. Unlike the dual Higgs model with dual Dirac strings [11,12]
the mass of a dual--vector field $M_C$ is not proportional to the order
parameter $<\bar{\chi}\chi>$ and does not vanish in the limit
$<\bar{\chi}\chi> \to 0$. This is seen from the mass formula [4]
\begin{eqnarray}\label{label4.2}
M_{\sigma}\,( 8\,M^2_C + 3\,M^2_{\sigma}) = - 56 g^2 <\bar{\chi}\chi>,
\end{eqnarray}
which can be derived from Eq.(\ref{label1.2}) and the gap--equation
Eq.(\ref{label1.4}). Thus,  in the MNJL model the dual--vector field does
not need a Goldston boson as a longitudinal component. This distinguishes
the transition to the non--perturbative superconducting phase in the NMJL
and the dual Higgs model. Indeed, in the MNJL model this transition does
not accompany the appearance of Goldston bosons. The former is rather
natural, since the starting $U(1)$ magnetic symmetry in the MNJL model is
global and unbroken in the non--perturbative superconducting phase. Recall
that in the dual Higgs model the magnetic $U(1)$ symmetry is local and
becomes spontaneously broken in the superconducting phase.

Due to the independence of the mass of the dual--vector field on the
monopole condensate the string tension $\sigma_0$ calculated in the
tree--approximation does not depend on the monopole condensate too. The
mass of the Higgs field $M_{\sigma}$ replaced the cut--off
$\Lambda_{\perp}$, i.e. $\Lambda_{\perp} = M_{\sigma}$. The dependence on
the magnetic monopole condensate appears by virtue of the contributions of
the quantum field fluctuations of the dual--vector $C_{\mu}$ and the scalar
$\sigma$ fields around the shape of the Abrikosov flux line. Very similar
to the dual Higgs model with dual Dirac strings the quantum field
fluctuations increase the value of the string tension. This implies that
for the consistent investigation of the superconducting mechanism of the
quark confinement within the dynamics of magnetic monopoles and dual Dirac
strings one cannot deal with a classical level only and quantum
contributions should be taken into account. The string shape fluctuations
of dual Dirac strings induce a Coulomb--like universal contribution
calculated for openned strings by L\"uscher {\it et al.} [8] and for closed
strings by Faber {\it et al.} [7].


\newpage

\begin{thebibliography}{9}
\bibitem{[1]}
P. Becher and H. Joos, Z. Phys. C15 (1982) 343;
V. Singh, D. Browne and R. Haymaker, Phys. Rev. D47 (1993) 1715
\bibitem{[2]}
M. Faber, A. N. Ivanov, W. Kainz and N. I. Troitskaya,
Z. Phys. C74 (1997) 721.
\bibitem{[3]}
M. Faber, A. N. Ivanov, W. Kainz and N. I. Troitskaya,
Phys. Lett. B386 (1996) 198.
\bibitem{[4]}
M. Faber, A. N. Ivanov, A. M\"uller, N. I. Troitskaya and M. Zach,
"Quantum and string shape fluctuations in the dual Monopole
Nambu--Jona--Lasinio model", hep--th/9805166, to appear
in European Physical Journal C.
\bibitem{[5]}
M. Faber, A. N. Ivanov, W. Kainz and N. I. Troitskaya,
Nucl. Phys. B475 (1996) 73.
\bibitem{[6]}
M. Faber, A. N. Ivanov, N. I. Troitskaya and M. Zach,
Phys. Lett. B400 (1997) 145.
\bibitem{[7]}
M. Faber, A. N. Ivanov, N. I. Troitskaya and M. Zach,
Phys. Lett. B409 (1997) 331.
\bibitem{[8]}
M. L\"uscher, K. Symanzik and P. Weisz,
Nucl.Phys. B173 (1980) 365;
M. L\"uscher,
Nucl. Phys. B180 (1981) 317.
\end{thebibliography}
\end{document}